\documentclass[12pt]{iopart}
\usepackage{bm,amssymb,graphicx}
\usepackage{iopams}
\usepackage{color}
\newcommand{\medio}[1]{\left\langle #1\right\rangle}

\begin{document}

\title[Time evolution of the HCS]{Generalized time evolution of the homogeneous cooling state of a granular gas with positive and negative coefficient of normal restitution}
\author{Nagi Khalil}
\address{IFISC (CSIC-UIB), Instituto de F\'isica Interdisciplinar y Sistemas Complejos, Campus Universitat de les Illes Balears, 
E-07122, Palma de Mallorca, Spain.} \ead{nagi@ifisc.uib-csic.es}

\begin{abstract}
  The homogeneous cooling state (HCS) of a granular gas described by the inelastic Boltzmann equation is reconsidered. As usual, particles are taken as inelastic hard disks or spheres, but now the coefficient of normal restitution $\alpha$ is allowed to take negative values $\alpha\in[-1,1]$, a simple way of modeling more complicated inelastic interactions. The distribution function of the HCS is studied at the long--time limit, as well as for intermediate times. At the long--time limit, the relevant information of the HCS is given by a scaling distribution function $\phi_s(c)$, where the time dependence occurs through a dimensionless velocity $c$. For $\alpha\gtrsim -0.75$, $\phi_s$ remains close to the gaussian distribution in the thermal region, its cumulants and exponential tails being well described by the first Sonine approximation. On the contrary, for $\alpha\lesssim -0.75$, the distribution function becomes multimodal, its maxima located at $c\ne 0$, and its observable tails algebraic. The latter is a consequence of an unbalanced relaxation--dissipation competition, and is analytically demonstrated for $\alpha\simeq -1$ thanks to a reduction of the Boltzmann equation to a Fokker--Plank--like equation. Finally, a generalized scaling solution to the Boltzmann equation is also found $\phi(c,\beta)$. Apart from the time dependence occurring through the dimensionless velocity, $\phi(c,\beta)$ depends on time through a new parameter $\beta$ measuring the departure of the HCS from its long--time limit. It is shown that $\phi(c,\beta)$ describes the time evolution of the HCS for almost all times. The relevance of the new scaling is also discussed. 
\end{abstract}



\maketitle

\section{Introduction \label{sec:1}}

The homogeneous cooling state (HCS) of a granular gas is one of the simplest states of an ensemble of grains that move freely between inelastic collisions. It describes a situation where the system remains spatially homogeneous and its granular temperature, proportional to its total kinetic energy, decreases monotonically in time. Moreover, if grains are modeled as smooth hard spheres with a constant and positive coefficient of normal restitution $\alpha$, for the long--time limit the HCS has two important properties. First, the granular temperature decays in time as $\sim 1/t^2$ according to Haff's law \cite{ha83}, and second, the distribution function depends on time only through its dependence on the granular temperature, its scaling form being close to the gaussian distribution in the thermal region and has exponential tails \cite{brrucu96,noer98,huorbr00,gonoba03,brpo06}. In addition, the time dependence of the scaling distribution occurs only throught a dimensionless velocity. Similar features are found if grains have different mechanical properties \cite{mapi99,gadu99}, are nonspherical \cite{vita12,rualweluhi16}, or even if the model includes other details of the grains, such as roughness/rotations \cite{luhumczi98,krbrpozi09,sakrsa11,vesakr14}, or velocity dependent collision coefficients \cite{brpo00,bobr09,dubopubr13}. 

The first objective of the present work is to complement the existence studies of the HCS by considering negative values of the coefficient of normal restitution ($\alpha<0$). The extension has two complementary motivations. On the one hand, some years ago numerical studies \cite{sabohabr10,mukrpo12} showed that two particles can collide with an effective negative $\alpha$. A natural question then concerns the effect of $\alpha$ being negative on the global dynamics of the system. The actual value of the coefficient of restitution depends in general on the exact geometry of the collision, however as a first approximation we can take it as a constant and consider the collision rule for hard spheres with $\alpha\in[-1,1]$, a model that can be theoretically studied by means of the kinetic Boltzmann equation and numerically solved with direct simulation Monte Carlo (DSMC). On the other hand, the new possibility opens up the door for a new phenomenology. Namely, one of the main features of granular matter is the indissolubly coupling between dissipation and dynamics: whenever there is a collision, and hence an eventual approach to equilibrium, there is a dissipation of energy, that is the system is also driven by itself out of equilibrium. In the case of hard spheres, when collisions become more and more elastic, $\alpha\to 1^-$, the relaxation dominates the dissipation, in the sense that the scaling distribution function of the HCS becomes close to the maxwellian. The situation is completely different in the new elastic limit, $\alpha\to-1^+$, since now the equilibration mechanism of collisions as well as the dissipation become small (for $\alpha=-1$ there is no collision at all, i.e. the ideal gas limit). 

Despite its simplicity, the HCS is one of the most important states of a granular gas, since it plays the same role the equilibrium distribution plays for molecular gases. It turns out that a general hydrodynamics description for the grains can be derived, in the context of the Kinetic Theory, by using the long--time limit of the distribution function of the HCS as a reference state \cite{brdukisa98,sego98}. However, the aforementioned hydrodynamics has several limitations, with at least two different origins. First, the dynamics of the system has to be such that the long--time limit of the HCS has to be reached by the system in a time scale much smaller than the hydrodynamic ones. This puts some limits to the values of the gradients and the dissipation. In fact, some authors argue that the usual hydrodynamics only applies in the elastic limit $\alpha\simeq 1$ \cite{sego98}, while others claim that it also works for a wider range of dissipation \cite{brdukisa98}. Second, the usual hydrodynamics may also fail when describing some steady states of the grains where there is a direct coupling between gradients and dissipation, like in the uniform shear flow. In this cases, it is still possible to derive accurate closed hydrodynamic equations if, instead of the HCS, the appropriate reference state is identified \cite{lu06,ga06}. Interestingly, even though the simplest cases now correspond to homogeneous and steady situations (homogeneous and steady granular temperature), consistency requires that the new reference distribution function does depend on time. The very same situation is found when the system is globally driven by a thermostat \cite{gachave13,khga13} and/or a plate \cite{brbumaga15}. 

Coming back to the free--cooling case, if one makes an appropriate change of the time variable \cite{lu01}, say $\tau\sim \ln t$, the dynamics changes so that particles become accelerated while collisions are not affected. At the long--time limit, the dissipation of energy due the inelastic collisions are compensated for by the acceleration, and the HCS becomes a steady state. The change of the time variable maps, in a reversible way, the original system into a granular gas in contact with an effective thermostat \cite{lu01,brrumo04,lubrdu02,dubrlu02}. Now we come across an apparent contradiction: the reference state of a gas in contact with a thermostat should be time dependent for consistency \cite{gamatr13}, but the long--time limit of the HCS is a steady state in the new representation. The second objective of the present work is to unveil the contradiction by reconsidering the time evolution of the HCS in the steady--state representation. 

The work is organized as follows. Next section presents the model, its kinetic description, and the steady--state representation of the HCS. Section \ref{sec:3} contains the theoretical and numerical results for the long--time limit for $\alpha\in[-1,1]$. The limit $\alpha\to -1$ is studied with some details after a reduction of the inelastic Boltzmann equation to a Fokker--Plank type equation. The time evolution of the HCS is the content of Sec. \ref{sec:4}. Finally, Sec. \ref{sec:5} is devoted to the conclusions. 

\section{Model, kinetic description, and steady--state representation \label{sec:2}}
\subsection{The model}
The system is modeled as an ensemble of $N$ hard spheres in $d$ dimensions, inside a square box of volume $L^d$ with periodic boundary conditions. Particles have mass $m$, diameter $\sigma$, and move freely between collisions. If two particles with velocities $\mathbf{v}_1$ and $\mathbf{v}_2$ collide, their new velocities $\mathbf{v}_1'$ and $\mathbf{v}_2'$ are
\begin{eqnarray}
  \label{eq:1}
    && \mathbf{v}_1'=\mathbf{v}_1-\frac{1+\alpha}{2} [(\mathbf{v}_1-\mathbf{v}_2)\cdot \hat{\boldsymbol{\sigma}}]\hat{\boldsymbol{\sigma}}, \\
    && \mathbf{v}_2'=\mathbf{v}_2+\frac{1+\alpha}{2} [(\mathbf{v}_1-\mathbf{v}_2)\cdot \hat{\boldsymbol{\sigma}}]\hat{\boldsymbol{\sigma}},
\end{eqnarray}
with $\hat{\boldsymbol{\sigma}}$ being a unit vector pointing from second to first particle at contact. The coefficient of normal restitution $\alpha$ is a number in $[-1,1]$ and characterizes the amount of energy $\Delta E$ dissipated in the collision as
\begin{equation}
  \label{eq:2}
  \Delta E=-\frac{m}{4}(1-\alpha^2) [(\mathbf{v}_1-\mathbf{v}_2)\cdot \hat{\boldsymbol{\sigma}}]^2.
\end{equation}

Observe that for the same precollisional velocities the energy dissipation is the same regardless the sign of $\alpha$, while the postcollisional velocities are different in both cases. In fact, the two signs of $\alpha$ represent two different physical situations: while for $\alpha>0$ we have collisions of smooth hard spheres, for $\alpha<0$ Eq. (\ref{eq:2}) is a simplification (i.e. gives the asymptotic velocities) of a more complicated interaction that may involve overlaps of particles and/or include a rotation of the contact plane of the two spheres \cite{sabohabr10,mukrpo12}. More clearly, for $\alpha=1$ particles collide elastically while for $\alpha=-1$ there is no collision at all (ideal gas). One still can forget about the physical meaning of $\alpha<0$, and consider the collision rule as part of a new model of a granular gas that becomes close to the ideal gas at some limit, a limit that has it intrinsic interest. 

\subsection{Boltzmann kinetic equation}
We consider situations where the granular gas is spatially homogeneous and dilute enough. In this cases, the distribution function of the system $f(\mathbf{v},t)$, defined so that $f(\mathbf{v},t) d\mathbf v$ is the mean density of particles with velocity around $\mathbf v$ at time $t$, verifies the kinetic Boltzmann equation, that for our model takes the form 
\begin{equation}
  \label{eq:3}
  \frac{\partial }{\partial t}f(\mathbf v,t)=\sigma^{d-1}J[\mathbf v|f],
\end{equation}
with $J$ being the collision operator 
\begin{eqnarray}
  \label{eq:4}
\nonumber    J[\mathbf v_1|f]= \int d\mathbf v_2 \int d\hat{\boldsymbol{\sigma}}\ &&  \Theta [(\mathbf{v}_1-\mathbf{v}_2)\cdot \hat{\boldsymbol{\sigma}}](\mathbf{v}_1-\mathbf{v}_2)\cdot \hat{\boldsymbol{\sigma}} \\ && \times \left[\alpha^{-2} f(\mathbf v_1^*,t)f(\mathbf v_2^*,t)-f(\mathbf v_1,t)f(\mathbf v_2,t)\right].
\end{eqnarray}
The new velocities $\mathbf v_1^*$ and $\mathbf v_2^*$ are the precollisional ones, obtaining by inverting Eq. (\ref{eq:1})
\begin{eqnarray}
  \label{eq:5}
    && \mathbf{v}_1^*=\mathbf{v}_1-\frac{1+\alpha}{2\alpha} [(\mathbf{v}_1-\mathbf{v}_2)\cdot \hat{\boldsymbol{\sigma}}]\hat{\boldsymbol{\sigma}}, \\
    && \mathbf{v}_2^*=\mathbf{v}_2+\frac{1+\alpha}{2\alpha} [(\mathbf{v}_1-\mathbf{v}_2)\cdot \hat{\boldsymbol{\sigma}}]\hat{\boldsymbol{\sigma}}.
\end{eqnarray}

By definition, the distribution function is normalized to the density of particles,
\begin{equation}
  \label{eq:6}
  \frac{N}{L^d}=\int d\mathbf v f(\mathbf v,t),
\end{equation}
If we define the mean of any quantity $A(\mathbf v)$ as 
\begin{equation}
  \label{eq:7}
  \medio{A}=\frac{L^d}{N} \int d\mathbf v \ A(\mathbf v) f(\mathbf v,t),
\end{equation}
then the granular temperature is $\frac{m}{d}\medio{v^2}$ and the first two cumulants read
\begin{eqnarray}
  \label{eq:8}
  && a_2=\frac{d}{d+2}\frac{\medio{v^4}}{\medio{v^2}^2}-1, \\
  \label{eq:9}
  && a_{3,s}=-\frac{d^2}{(d+2)(d+4)}\frac{\medio{v^6}}{\medio{v^2}^3}+\frac{3d}{d+2}\frac{\medio{v^4}}{\medio{v^2}^2}-2.
\end{eqnarray}
Very often, the latter quantities are used to characterize the HCS instead of the distribution function itself.

\subsection{Steady--state representation}

Since collisions dissipate energy for $\alpha\in(-1,1)$, the granular temperature is always a decreasing function of time. However, the change of the time variable introduced in \cite{lu01} and further analyzed in \cite{brrumo04} enables the system to reach a steady state. The new time scale does not change the dynamics of the system, but rather represents an useful way of observing it. 

Let $\nu_0$ and $t_0$ be arbitrary positive constants, with dimensions of frequency and time, respectively. We define a new time variable $\tau$ as
\begin{eqnarray}
  \label{eq:10}
  \tau=\frac{1}{\nu_0}\ln \frac{t}{t_0}.
\end{eqnarray}
Then, if $\mathbf r$ denotes the position of a particle, its velocity defined in terms of the new time is $\mathbf w=\frac{d}{d\tau}\mathbf r$ which gives 
\begin{equation}
  \label{eq:11}
  \mathbf w(\tau)=\nu_0 t\mathbf v(t),
\end{equation}
that is, particles are accelerated between collisions. Moreover, since collisions are instantaneous, the same rules (\ref{eq:1}) and (\ref{eq:5}) apply for $\mathbf w$, after replacing $\mathbf v$ by $\mathbf w$. 

At the kinetic level, it is convenient to consider a new distribution function $g(\mathbf w,\tau)$ defined as
\begin{equation}
  \label{eq:12}
  g(\mathbf w,\tau)=\left(\nu_0 t\right)^{-d}f(\mathbf v,t)
\end{equation}
which has the same normalization of $f$. Taking into account Eq. (\ref{eq:3}) for $f$, we get the following equation for $g$:
\begin{equation}
  \label{eq:13}
  \left[\frac{\partial}{\partial \tau}+\nu_0\frac{\partial}{\partial \mathbf w}\cdot \mathbf w\right]g(\mathbf w,\tau)=\sigma^{d-1}J[\mathbf w|g],
\end{equation}
where the collision operator $J$ is defined by Eq. (\ref{eq:4}). The fundamental difference of the present equation with respect to Eq. (\ref{eq:3}) is the presence of a new term in the l.h.s. As already mentions in Refs. \cite{lu01,brrumo04}, the new term acts as a thermostat, injecting energy into the system, and allowing it to reach a steady state, as we analyze in the next section. 

\section{Long--time limit  \label{sec:3}}
Equation (\ref{eq:13}) has a steady--state solution $g_s(\mathbf w,\tau)$ describing the long--time limit of the distribution function of the HCS. For $\alpha>0$, the research on the distribution function has focused mainly on two aspects: their cumulants (providing information of the thermal region) and the tails, see for instance \cite{brrucu96,noer98,huorbr00,gonoba03,brpo06,brrumo04}. In this section, we extend the study to negative values of the coefficient of normal restitution.

\subsection{Scaling solution} 

The steady--state solution of Eq. (\ref{eq:13}) admits the following scaling form
\begin{eqnarray}
  \label{eq:14}
  g_s(\mathbf c,\tau)=\frac{N}{L^d}w_0^{-d}\phi_s(\mathbf c),
\end{eqnarray}
where $w_0$ and $\mathbf c$ are defined in general as 
\begin{eqnarray}
  \label{eq:15}
    && w_0=\sqrt{\frac{2T}{m}},\\
    && \mathbf c=\frac{\mathbf w}{w_0}.
\end{eqnarray}
with $T$ being the temperature associated to $g$ (not with $f$) as
\begin{equation}
  \label{eq:16}
  T(\tau)=\frac{m}{d}\frac{L^d}{N}\int d\mathbf w\ w^2 g(\mathbf w,\tau).
\end{equation}
Using relations (\ref{eq:11}) and (\ref{eq:12}), it is easily seen that the granular temperature is $(\nu_0t)^{-2}T$. In Eq. (\ref{eq:14}), it is $T=T_s$. Multiplying Eq. (\ref{eq:13}) by $w^2$, integrating over velocities, and after some algebra, we arrive at an equation for $T_s$. The solution reads 
\begin{equation}
  \label{eq:17}
  T_s=\frac{m}{2}\left(\frac{2L^d\nu_0}{N\sigma^{d-1} \zeta^*_s}\right)^2,
\end{equation}
where $\zeta^*_s$ is a dimensionless cooling rate given by
\begin{equation}
  \label{eq:18}
  \zeta^*_s(\alpha)=\frac{(1-\alpha^2)\pi^{\frac{d-1}{2}}}{2d\Gamma\!\left(\frac{d+3}{2}\right)} \int d\mathbf{c_1}d\mathbf{c_2}\ |\mathbf c_1-\mathbf c_2|^3 \phi_s(\mathbf c_1)\phi_s(\mathbf c_2).  
\end{equation}
For a given value of $\nu_0$, the steady--state temperature is a function of the coefficient of normal restitution through $\zeta^*_s$, not only through the factor $(1-\alpha^2)$ at Eq. (\ref{eq:18}), but also through an implicit dependence on $\phi_s$. Hence, to complete the computation of $T_s$, we need $\phi_s$. 

The first moments of the scaling function $\phi_s(\mathbf c)$ are directly given by relation (\ref{eq:14}). Namely, it is normalized to one, has zero mean, and its second moment is $d/2$. Its corresponding equation can be deduced by replacing the scaling form of Eq. (\ref{eq:14}) into Eq. (\ref{eq:13}) and dropping the time derivative
\begin{equation}
  \label{eq:19}
  \frac{\zeta^*_s}{2}\frac{\partial}{\partial \mathbf c}\cdot \left[\mathbf c \phi_s(\mathbf c)\right]=J[\mathbf c|\phi_s].
\end{equation}
This new equation has to be solved together with Eq. (\ref{eq:18}), with the additional knowledge of its first moments. Note that we obtain the same equation if we impose the scaling form $f=N/L^d(\nu_0t)^d w_0^{-d}\phi_s(\mathbf c)$ to the original Boltzmann equation (\ref{eq:3}), see \cite{noer98} for further details. 

\subsection{Cumulants and tails}
Equation (\ref{eq:19}) has an isotropic solution $\phi_s(c)$ that can be expanded in terms of the Sonine polynomials \cite{noer98} $S_n(c^2)$ as 
\begin{equation}
  \label{eq:20}
  \phi_s(\mathbf c)=\pi^{-\frac{d}{2}}e^{-c^2}\left[1+\sum_{n=2}^\infty a_{n,s}S_n(c^2)\right],
\end{equation}
Although the expansion breaks down for moderate dissipation \cite{brpo06}, useful information is gained if only the leading term is retained, namely $S_2(c^2)=c^4/2-(d+2)c^2/2 +d(d+2)/8$. The coefficient $a_{2,s}(\alpha)$ coincides with the first cumulant of the distribution, defined if Eq. (\ref{eq:8}). 

At the leading approximation of $\phi_s( c)$, a closed equation for $a_{2,s}$ can be obtained from Eq. (\ref{eq:19}). In addition, if terms proportional to $a_{2,s}^2$ are neglected, we end up with the following approximate expressions \cite{brrumo04,noer98}  for the first cumulant 
\begin{equation}
  \label{eq:21}
  a_{2,s}(\alpha)\simeq\frac{16(1-\alpha)(1-2\alpha^2)}{9+24d+(8d-41)\alpha+30\alpha^2(1-\alpha)}
\end{equation}
and for the scaled cooling rate
\begin{equation}
  \label{eq:22}
  \zeta^*_s(\alpha)\simeq \frac{\sqrt{2}(1-\alpha^2)\pi^{\frac{d-1}{2}}}{d\Gamma\!\left(\frac{d}{2}\right)} \left[1+\frac{3}{16}a_{2,s}(\alpha)\right].
\end{equation}
An approximate expression for $T_s$ is now obtained by replacing Eq. (\ref{eq:22}) for $\zeta^*_s$ into Eq. (\ref{eq:17}). See Refs. \cite{mosa00} and \cite{samo09} for more accurate approaches.

\begin{figure}[!h]
  \centering
 \includegraphics[width=.475\textwidth]{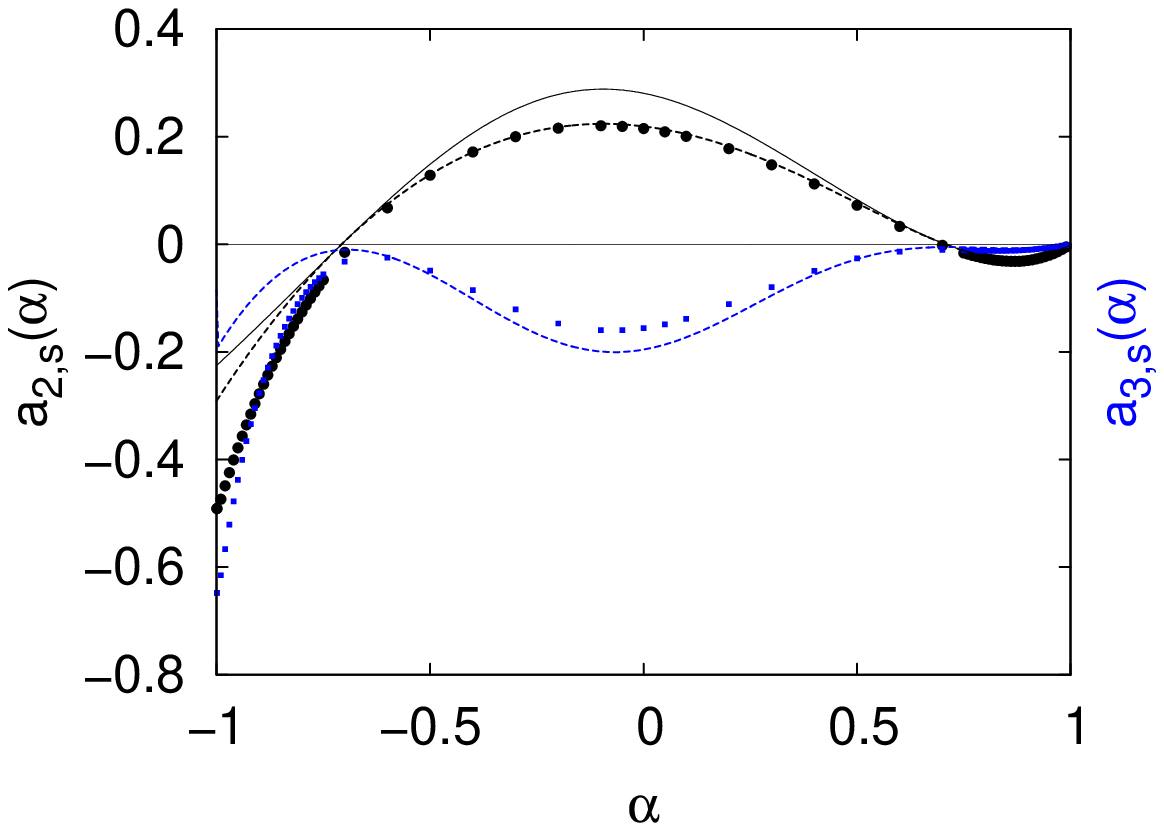}
 \includegraphics[width=.475\textwidth]{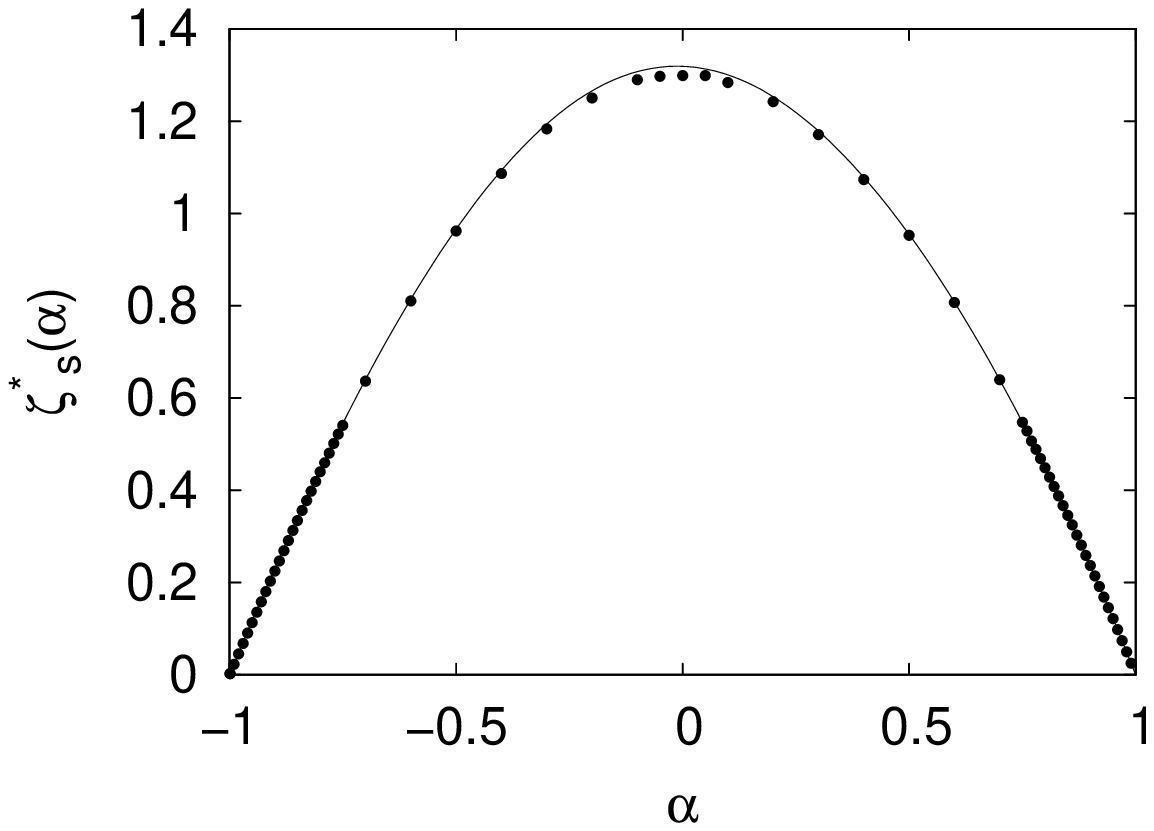}
 \caption{Results for a two dimensional system ($d=2$). Left: steady--state value from the theoretical prediction of Eq. (\ref{eq:21}) (solid line), form Ref. \cite{samo09} (dashed lines), and from simulations (circles for $a_2$ and squares for $a_{3,s}$), as a function of the coefficient of restitution $\alpha$. Right: steady--state values from the theory (line) and simulation (symbols) for the scaled cooling rate $\zeta_s^*$ given by Eq. (\ref{eq:18}) as a function of $\alpha$.}
 \label{fig:1}
\end{figure}

In Fig. \ref{fig:1}, the theoretical predictions of Eqs. (\ref{eq:21}) and (\ref{eq:22}) for $a_{2,s}$ and $\zeta^*_s$ are compared against DSMC numerical results. Numerical results for the third cumulant $a_{3,s}$, as well as theoretical results for the second and third model from \cite{samo09}, are also provided. As it is apparent, the theoretical prediction for the cooling rate $\zeta^*_s$ is very good even for $\alpha<0$. In fact, the important dependence on $\alpha$ is given by the factor $(1-\alpha^2)$, meaning that the sign of the coefficient of restitution is almost irrelevant for it. For the first cumulant $a_{2,s}$, theory deviates from simulations moderately for $|\alpha|\lesssim 0.5$, and notably  for $\alpha\lesssim -1/\sqrt{2}$. The values of $a_{3,s}$ suggest that the origin of the discrepancy relies on the truncation of expansion at Eq. (\ref{eq:20}) rather than on the assumption of the smallness of $a_2$, since the bigger $|a_{3,s}|$ the bigger the difference between theory and simulations, in agreement with the conclusions in \cite{brpo06} and \cite{samo09}. 

\begin{figure}[!h]
 \centering
 \includegraphics[width=.475\textwidth]{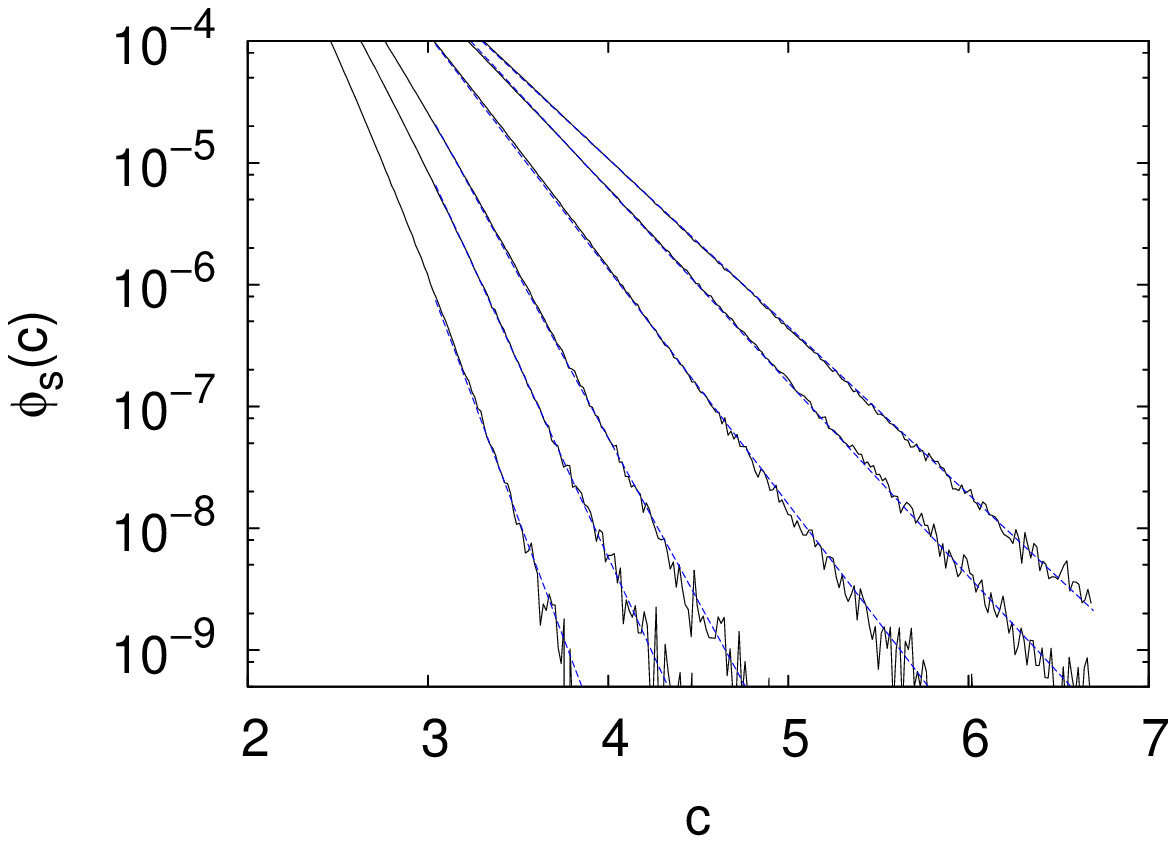}
 \includegraphics[width=.475\textwidth]{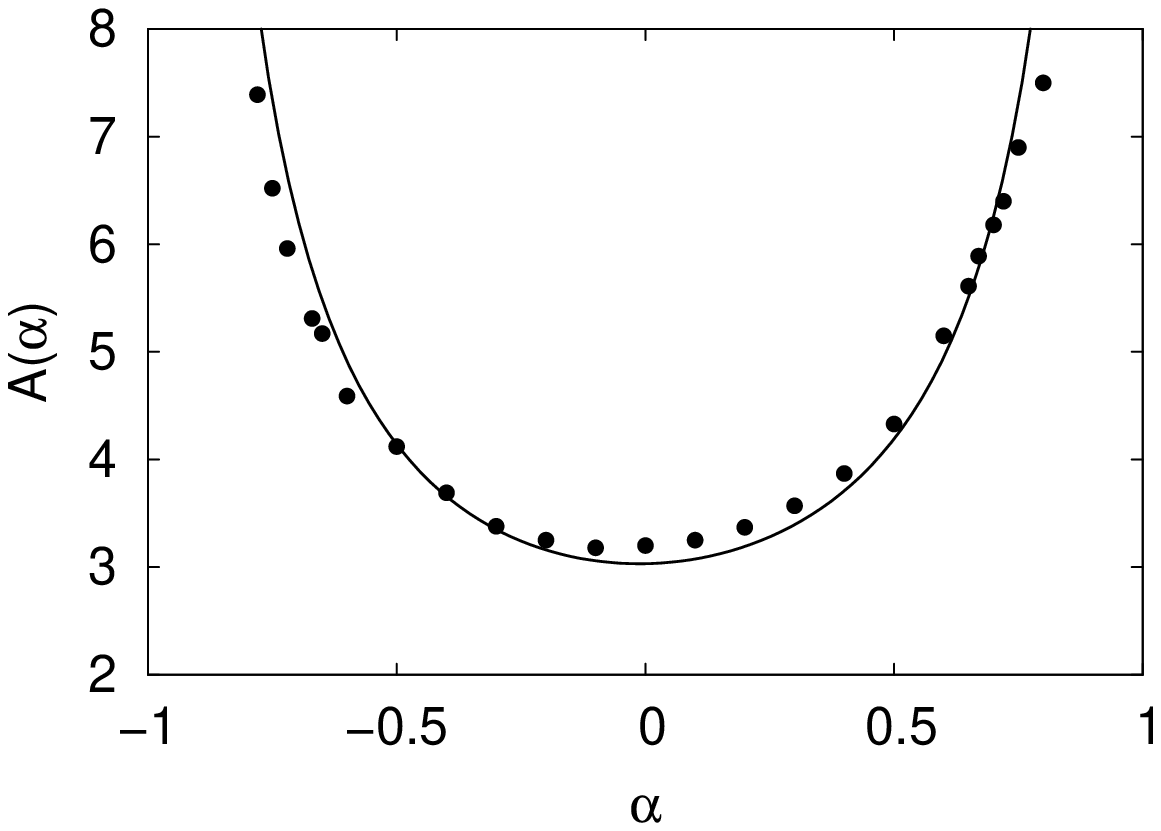}
 \caption{Left: exponential tails measured in simulations (black solid lines) and their respective fits (blue dashed lines) for $\alpha=-0.1;-0.4;-0.6;-0.75;-0.8;-0.85$ (from right to left). Right: numerical results obtained from the fits (dots) and the theoretical prediction (line) of Eq. (\ref{eq:23}) for the exponent $A(\alpha)$.}
 \label{fig:1a}
\end{figure}

The tails of the distribution function can be evaluated by following the arguments at \cite{noer98}. The distribution function has an exponential tail for $c\gg 1/(1-\alpha^2)$, that is $\phi_s\sim e^{-Ac}$ where the exponent is given, at the first Sonine approximation, by 
\begin{equation}
  \label{eq:23}
  A(\alpha)=\frac{\sqrt{2}d\Gamma\!\left(\frac{d}{2}\right)}{\Gamma\!\left(\frac{d+1}{2}\right)(1-\alpha^2)\left[1+\frac{3}{16}a_{2,s}(\alpha)\right]}.
\end{equation}
Figure \ref{fig:1a} shows a comparison of the latter equation and numerical simulations for a two--dimensional system. The exponent $A(\alpha)$ have been obtained by adjusting the numerical simulation in the region $c\lesssim 1/(1-\alpha^2)$. 

\subsection{Loss of unimodality}

\begin{figure}[!h]
 \centering
 \includegraphics[width=.475\textwidth]{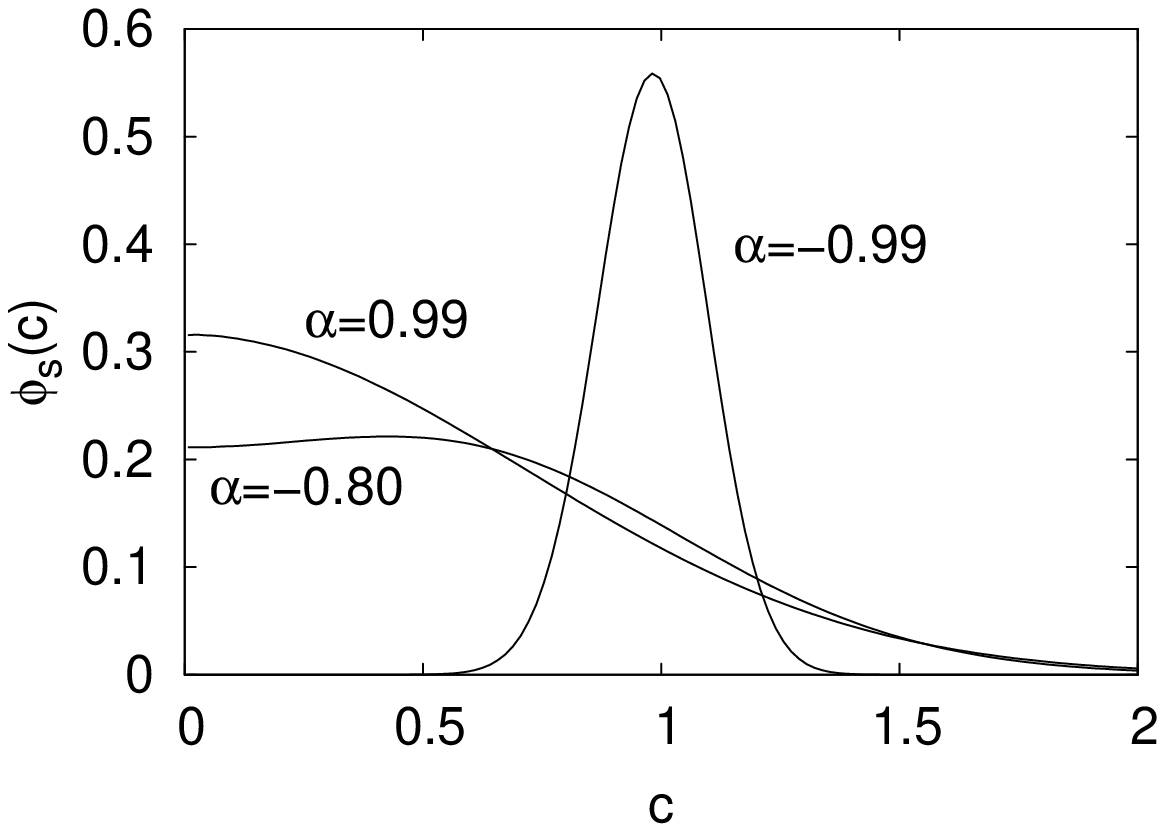}
 \includegraphics[width=.475\textwidth]{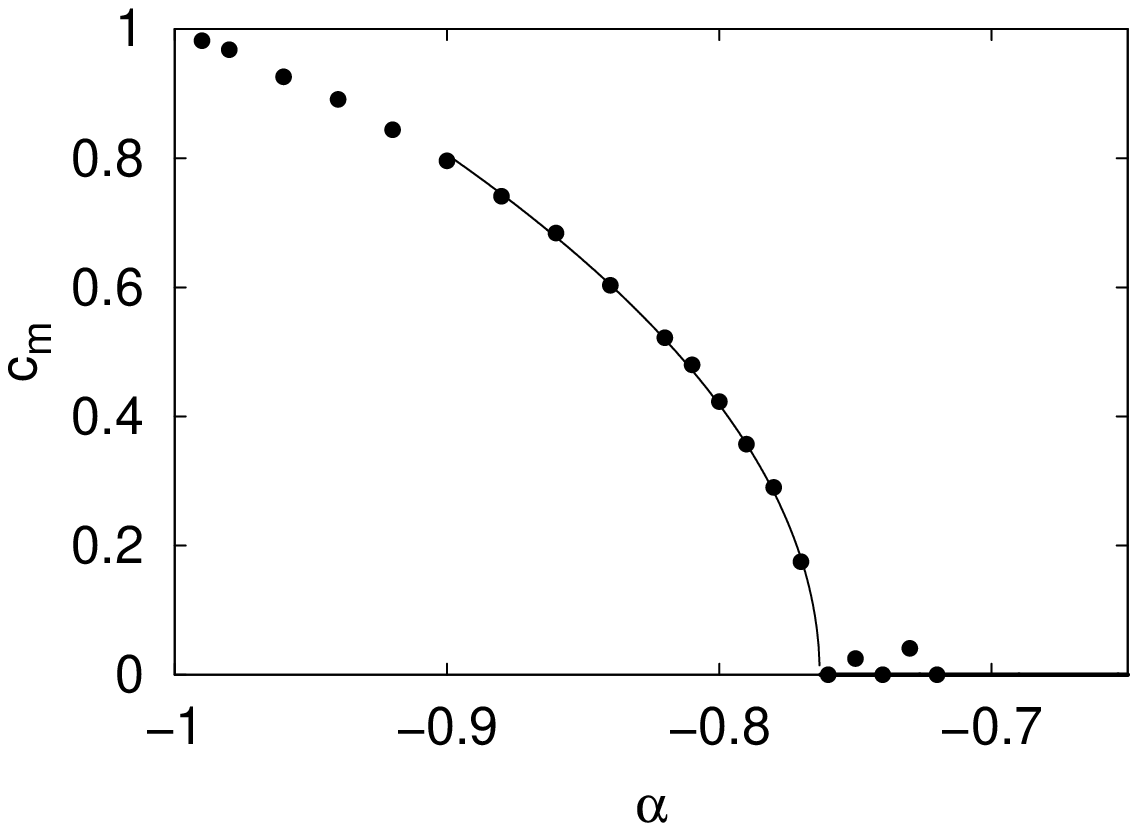}
 \caption{Left: Isotropic scaling distribution function $\phi_s(c)$ of the HCS at the long--time limit for a two dimensional system and for $\alpha=-0.99,-0.8,0.99$. Right: location of the maxima $c_m$ of $\phi_s(c)$: symbols are from the simulations and the line is a best fit to $\sim\sqrt{\alpha-\alpha_c}$.}
 \label{fig:2}
\end{figure}

An important conclusion is inferred from the theory and numerical results shown in Fig. \ref{fig:1}. Namely, the distribution functions for the HCS at the steady state for $\alpha$ and $-\alpha$ are different, specially if $|\alpha|\gtrsim 1/\sqrt{2}$ (the cumulants are quite different). The difference is dramatic for the two elastic limits: while for $\alpha\to 1$ the distribution function tends to the gaussian distribution (all cumulants go to zero), for $\alpha\to -1$ the asymptotic distribution is different from the gaussian (the cumulants are different from zero). This scenario is fully supported by the left plot of Fig. \ref{fig:2} where the scaling distribution $\phi(c)$ is plotted for three values of $\alpha$, one close to $1$ and two close to $-1$. 

Figure \ref{fig:2} shows a qualitative change of $\phi_s$ for $c\lesssim 2$ (thermal region) as $\alpha$ decreases. Namely, there is a critical value $\alpha_c$ of the coefficient of normal restitution behind which the distribution function becomes multimodal. More precisely, if $c_m$ is the absolute value of $\mathbf c$ where the maxima of $\phi_s(\mathbf c)$ occur, then the numerical simulations show that $c_m$ changes continuously as $c_m\sim \sqrt{\alpha_c-\alpha}$ for $\alpha_c\simeq -0.75$, see the right plot of Fig. \ref{fig:2}. 

The loss of unimodality for $\alpha<\alpha_c$ can be understand qualitatively as an unbalanced competition between the collision--inducing relaxation and the collision--inducing dissipation, as follows. Take Eq. (\ref{eq:19}) for $\mathbf c=0$, and write it as
\begin{equation}
\label{eq:24}
  J^+[0|\phi_s]-J^-[0|\phi_s]-\frac{d}{2}\zeta^*_s\phi_s(0)=0,
\end{equation}
where $J^\pm$ are the contributions of the collision operator that accounts for the gain ($+$) and loose ($-$) of particles with velocity $\mathbf c$ around zero. For $|\alpha|\simeq 1$ the term proportional to $\zeta^*_s$ can be dropped and the argumentation simplified. First, take as a reference $\alpha\simeq 1$: the mean number of collisions that produces a particle with velocity around the origin $J^+$ coincides with the mean number of collisions that moves a particle from the origin to any other place $J^-$. For this case, $J^+$ involves particles with a wide range of velocities. Now take $\alpha\simeq -1$, and assume that $J^-$ does not change too much with respect to the previous case, since it depends essentially on the collision frequency. Since $1+\alpha\sim 0$ and the distribution function decays to zero very fast for $c\gtrsim 3$, the only way a particle around the origin to be created is that it was originally nearby. Hence, the wide range of pairs of velocities of case $\alpha\simeq 1$ now reduces to a narrow region, that is the distribution function has to be bigger around the origin now. 

\subsection{The limit $\alpha\to -1^+$}

The previous qualitative argumentation about the loss of unimodality can be done formally for $\alpha\simeq -1$, which has its intrinsic interest. Taking advantage of the fact that $\epsilon \equiv (1+\alpha)/(2\alpha)$ is very small for $\alpha\to -1^+$, the collision operator $J$ defined in Eq. (\ref{eq:4}) can be simplified. The procedure is similar to that followed upon deriving the Fokker--Plank equation for a brownian particle from the Boltzmann--Lorentz equation \cite{brdusa99}. By expanding the distribution function appearing at $J$ up to order $\epsilon^2$, neglecting higher-order contributions and contributions form velocities such as $c |\epsilon| \gtrsim 1$, we get
\begin{equation}
  \label{eq:25}
  J=\epsilon J_1+\epsilon^2J_2+\mathcal O(\epsilon^4), 
\end{equation}
with 
\begin{eqnarray}
  \label{eqa:26}
    J_1[\mathbf c|\phi]=&&-\frac{\pi^{\frac{d-1}{2}}}{\Gamma\!\left(\frac{d+3}{2}\right)} \frac{\partial}{\partial \mathbf c}\cdot\left\{\mathbf I[\mathbf c|\phi] \phi(\mathbf c)\right\}, \\
    \mathbf I[\mathbf c|\phi]=&&\int d\mathbf c_1 |\mathbf c-\mathbf c_1|(\mathbf c-\mathbf c_1)\phi(\mathbf c_1),
\end{eqnarray}
and 
\begin{eqnarray}
  \label{eqa:27}
\nonumber     J_2[\mathbf c|\phi]=&&\frac{\pi^{\frac{d-1}{2}}}{2\Gamma\!\left(\frac{d+5}{2}\right)}\left\{\frac12 \frac{\partial}{\partial c_i}\frac{\partial}{\partial c_j}\left\{I_{ij}[\mathbf c|\phi] \phi(\mathbf c)\right\} \right. \\ && \qquad \left. -2(d+3) \frac{\partial}{\partial \mathbf c}\cdot\left\{\mathbf I[\mathbf c|\phi] \phi(\mathbf c)\right\}\right\}, \\
     I_{ij}[\mathbf c|\phi]=&&\int d\mathbf c_1 \left[3|\mathbf c-\mathbf c_1|(\mathbf c-\mathbf c_1)_i(\mathbf c-\mathbf c_1)_j+|\mathbf c-\mathbf c_1|^3\delta_{ij}\right]\phi(\mathbf c_1).
\end{eqnarray}

If the expansion is now used with Eq. (\ref{eq:19}), the equation for the steady--state solution $\phi_s$ of the HCS becomes of a Fokker--Plank type,
\begin{equation}
  \label{eq:28}
\frac{\partial}{\partial c_i}\frac{\partial}{\partial c_j}\left\{D_{ij}[\mathbf c|\phi_s]\phi_s(\mathbf c)\right\}+\frac{\partial}{\partial \mathbf c}\cdot\left\{\boldsymbol \mu_2[\mathbf c|\phi_s]\phi_s(\mathbf c)\right\}=0,
\end{equation}
with the diffusion and drift terms being functionals of $\phi_s$:
\begin{eqnarray}
  \label{eq:29}
    D_{ij}[\mathbf c|\phi]=&&\frac{\epsilon}{2(d+3)}I_{ij}[\mathbf c|\phi], \\
    \boldsymbol \mu_2[\mathbf c|\phi]=&&\frac1d \left[\int d\mathbf c_1 d\mathbf c_2|\mathbf c_1-\mathbf c_2|^3\phi(\mathbf c_1)\phi(\mathbf c_2)\right] \mathbf c -\mathbf I[\mathbf c|\phi].
\end{eqnarray}
Equation (\ref{eq:28}) has to be consistent with the known values of the fist moments of $\phi_s$, namely $\int d\mathbf c \ \phi_s=1$, $\int d\mathbf c \ \mathbf c \phi_s=1$, and $\int d\mathbf c \ c^2 \phi_s=1$. By taking moments of the equation, and assuming that the distribution function decays to zero fast enough for $c\gg 1$ (something to be proved bellow), the consistency is easily proved. Equations (\ref{eq:28}) and (\ref{eq:29}) imply that the limit $\epsilon\to 0$ is singular, that is, the Boltzmann equation for $\alpha=-1$ (ideal gas case) is different from $\alpha=-1^+$, since for the latter case the equation reduces to Eq. (\ref{eq:28}) with a vanishing diffusion term.

As an application of Eq. (\ref{eq:28}), we can infer that $c=0$ is a local minimum of $\phi_s(c)$. Putting $c=0$ into Eq. (\ref{eq:28}) and after some algebra we get 
\begin{eqnarray}
  \label{eq:30}
\nonumber    \frac{d^2\phi_s(0)}{dc^2}=&&-\frac{2d}{\epsilon}\int d\mathbf c_1 d\mathbf c_2 \bigg[|\mathbf c_1-\mathbf c_2|^3 \\ && \quad -\frac{d+1}{d}(c_1^2c_2+c_1c_2^2)\bigg]\phi_s(\mathbf c_1)\phi_s(\mathbf c_2), 
\end{eqnarray}
whose sign is not easy to obtain. However, if we replace, as a first approximation, $\phi_s$ by a gaussian, we get a positive second derivative. For a general function $\phi_s$, we can demonstrate that $\frac{d^2\phi_s(0)}{dc^2}>0$ for the one dimensional case ($d=1$), where, after some manipulations, we have
\begin{eqnarray}
  \label{eq:31}
  \nonumber \frac{d^2\phi_s(0)}{dc^2}=&&-\frac{2d}{\epsilon}\int_0^\infty dc_1 \int_0^\infty d c_2\big[|c_1-c_2|^3 \\ &&  +|c_1-c_2|^2(c_1+c_2)\big]\phi_s(c_1)\phi_s(c_2)>0.
\end{eqnarray}
Another way of realizing that $c=0$ is a local minimum is by considering the extreme elastic limit ($\alpha=-1^+$). Putting $\epsilon=0$ in Eq. (\ref{eq:28}), we have $\frac{\partial}{\partial \mathbf c}\cdot\left\{\boldsymbol \mu_2[\mathbf c|\phi_s]\phi_s(\mathbf c)\right\}=0$ which reduces, by isotropic considerations, to $\boldsymbol \mu_2[\mathbf 0|\phi_s]\phi_s(\mathbf 0)=\mathbf 0$ for $c=0$. Since $\boldsymbol \mu_2[\mathbf 0|\phi_s]\ne \mathbf 0$, it is $\phi_s(\mathbf 0)=0$. Since $\phi_s(\mathbf c)\ge 0$, we conclude that $c=0$ is a local minimum. This is exactly what the left plot of Fig. \ref{fig:2} shows for $\alpha=-0.99$. 

\begin{figure}[!h]
 \centering
 \includegraphics[width=.475\textwidth]{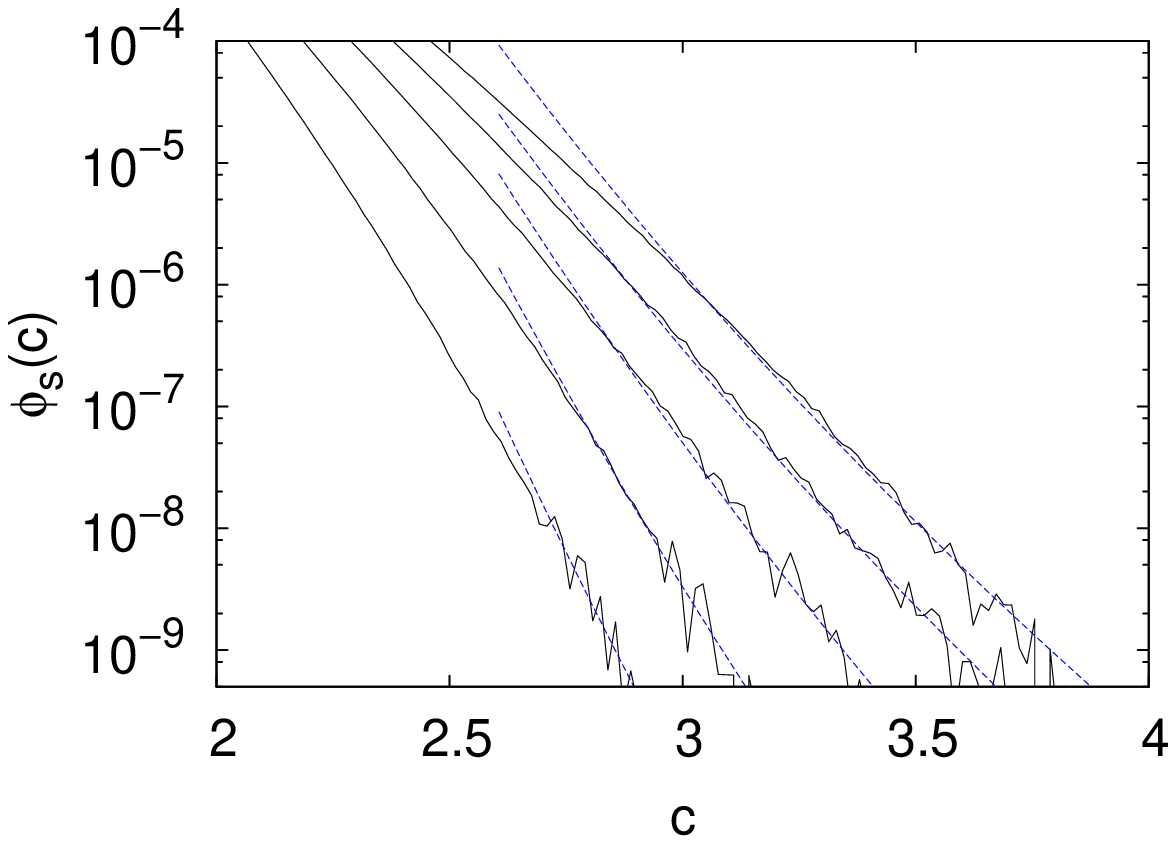}
 \includegraphics[width=.475\textwidth]{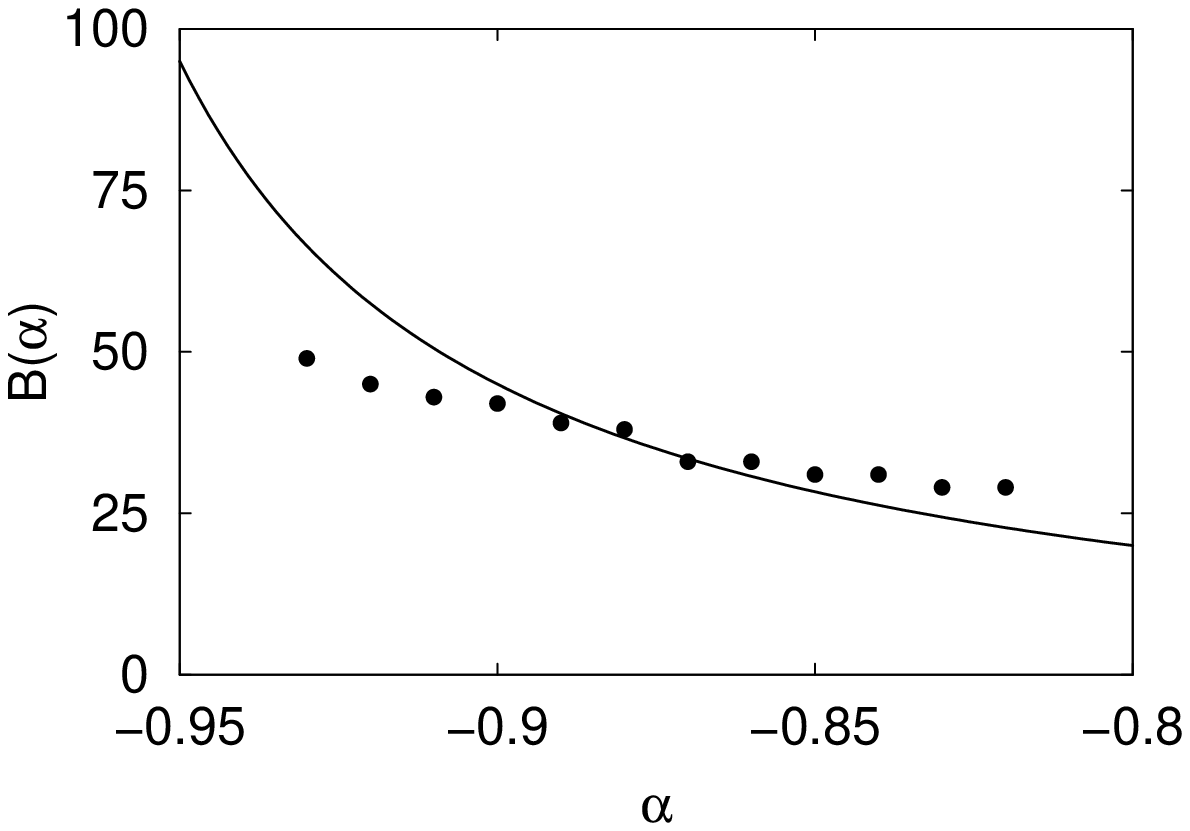}
 \caption{Left: Algebraic tails of $\phi_s(c)$ obtained from the simulations (black solid lines) and their best fits (blue dashed lines) for $\alpha=-0.85;-0.87;-0.89;-0.91;-0.93$ (from right to left). Right: exponent $B(\alpha)$ obtained from the simulations (dost) and its theoretical expression (line) given by Eq. (\ref{eq:34}).}
 \label{fig:2a}
\end{figure}

As another application, we also compute the tails of $\phi_s$. For that purpose, it is convenient to rewrite Eq. (\ref{eq:28}) as
\begin{equation}
  \label{eq:32}
  \frac{\partial}{\partial c_j}\left\{D_{ij}[\mathbf c|\phi_s]\phi_s(\mathbf c)\right\}+\frac{\partial}{\partial \mathbf c}\cdot\left\{\mu_i[\mathbf c|\phi_s]\phi_s(\mathbf c)\right\}=a_i,
\end{equation}
where $\mathbf a$ as constant vector. Since we seek for an isotropic solution $\phi_s(c)$, the only possibility is $a_i$ to be zero. If now we take $c\gg 1$, then the diffusion and drift terms simplifies, and the resulting equation becomes
\begin{equation}
  \label{eq:33}
  \frac{2\epsilon}{d+3}c^3\frac{d\phi_s(c)}{dc}=c^2\phi(c),
\end{equation}
giving rise to algebraic tails:
\begin{eqnarray}
  \label{eq:34}
    && \phi_s(c)\sim c^{-B(\alpha)}, \quad \alpha\sim -1^+, \\ 
    && B(\alpha)=\frac{(d+3)|\alpha|}{1+\alpha}.
\end{eqnarray}
Since we assumed $c\ll 1/|\epsilon|$ in the derivation of Eq. (\ref{eq:28}) (see just after Eq. (\ref{eq:25})), Eq. (\ref{eq:34}) is not in contradiction with the exponential tails of $\phi_s$ expected for $c\gg 1/|\epsilon|$. Figure \ref{fig:2a} compares the numerical simulations with Eq. (\ref{eq:34}) for the values of $\alpha$ for which the tails could be measured. 

\section{Generalized scaling solution \label{sec:4}}

In this section, we aim at describing the time evolution of the HCS at a wider time window, and in particular the approach towards its long--time limit. The starting point will be the time dependent equation for the distribution function in the steady--state representation (\ref{eq:13}). As already described in other works where a similar equation were analyzed \cite{gamatr13}, a consistent scaling solution to Eq. (\ref{eq:13}) requires a time dependence not only through the scaling velocity $\mathbf c$ but also through another dimensionless parameter $\beta$ as 
\begin{equation}
  \label{eq:35}
  g(\mathbf w,\tau)=\frac{N}{L^d} w_0^{-d}\phi(\mathbf c,\beta),
\end{equation}
where $w_0$ and $\mathbf c$ are defined in Eq. (\ref{eq:15}). If we substitute the scaling form  into Eq. (\ref{eq:13}), we get
\begin{equation}
  \label{eq:36}
  \left[-\frac{d\ln w_0}{d\tau}+\nu_0\right] \frac{\partial}{\partial \mathbf c}\cdot \left[\mathbf c \phi(\mathbf c,\beta)\right] +\frac{d\beta}{d\tau}\frac{\partial}{\partial \beta}\phi(\mathbf c,\beta)=nw_0\sigma^{d-1}J[\mathbf c,\beta|\phi]  
\end{equation}
which demonstrates the need of including $\beta(\tau)$ in order to cancel out the time dependence introduced by $w_0$. We have made explicit the new dependence of the collision operator on $\beta$ as $J[\mathbf c,\beta|\phi]\equiv J[\mathbf c|\phi]$.  A convenient election for $\beta$ is 
\begin{equation}
  \label{eq:37}
  \beta=\sqrt{\frac{T_s}{T}},
\end{equation}
which measures the deviation of the HCS ($\beta\ne 1$) form its long--time limit ($\beta= 1$). With this choose, an exact equation for $\beta(\tau)$ can be derived from Eq. (\ref{eq:35}) by multiplying it by $c^2$, integrating over $\mathbf c$, and imposing $\int d\mathbf c c^2 \phi=d/2$,
\begin{equation}
  \label{eq:38}
  \frac{d}{ds}\beta(s)+\mu_2[1|\phi]\beta(s) -\mu_2[\beta|\phi]=0.
\end{equation}
A dimensionless time scale $s$ has been introduced as
\begin{equation}
  \label{eq:39}
  ds=\frac{N}{L^d}\sigma^{d-1}\sqrt{\frac{2T_s}{m}}d\tau,
\end{equation}
and the new quantity $\mu_2$ is 
\begin{equation}
  \label{eq:40}
  \mu_2[\beta|\phi]=-\frac{1}{d}\int d\mathbf c\ c^2J[\mathbf c,\beta|\phi].
\end{equation}
Note that Eq. (\ref{eq:38}) is a first order differential equation where $\beta=1$ is a fixed point, as expected. 

The equation for $\phi$ can be now written in a consistent form as 
\begin{equation}
  \label{eq:41}
  \left\{\mu_2[\beta|\phi]-\beta \mu_2[1|\phi]\right\}\beta \frac{\partial}{\partial \beta}\phi(\mathbf c,\beta)+\mu_2[\beta|\phi]\frac{\partial}{\partial \mathbf c}\cdot \left[\mathbf c \phi(\mathbf c,\beta)\right] =J[\mathbf c,\beta|\phi],  
\end{equation}
and has to be solved with the knowledge of the first moments of $\phi$, namely $\int d\mathbf c\ \phi=1$, $\int d\mathbf c\ \mathbf c \phi=\mathbf 0$, and $\int d\mathbf c\ c^2 \phi=d/2$. Once it is solved, the time dependence of $\beta$ (and $T$) can be calculated by solving Eq. (\ref{eq:38}). Finally, this allows us to obtain the original distribution functions $g$ and $f$ for a wide range of times. Since $\mu_2[1|\phi]=\frac{1}{2}\zeta^*_s$, for $\beta=1$, Eq. (\ref{eq:19}) the steady--state distribution function $\phi_s$ is recovered.

\subsection{First Sonine approximation}

\begin{figure}[!h]
 \centering
 \includegraphics[width=.475\textwidth]{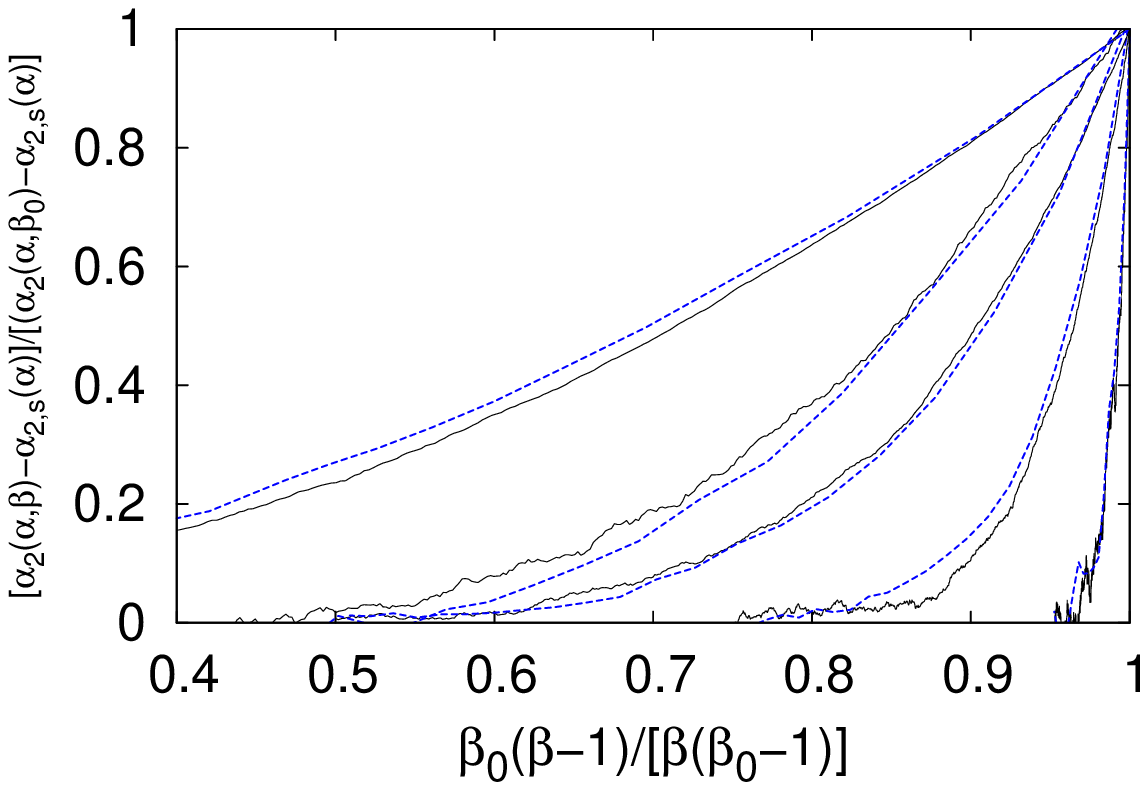}
 \includegraphics[width=.475\textwidth]{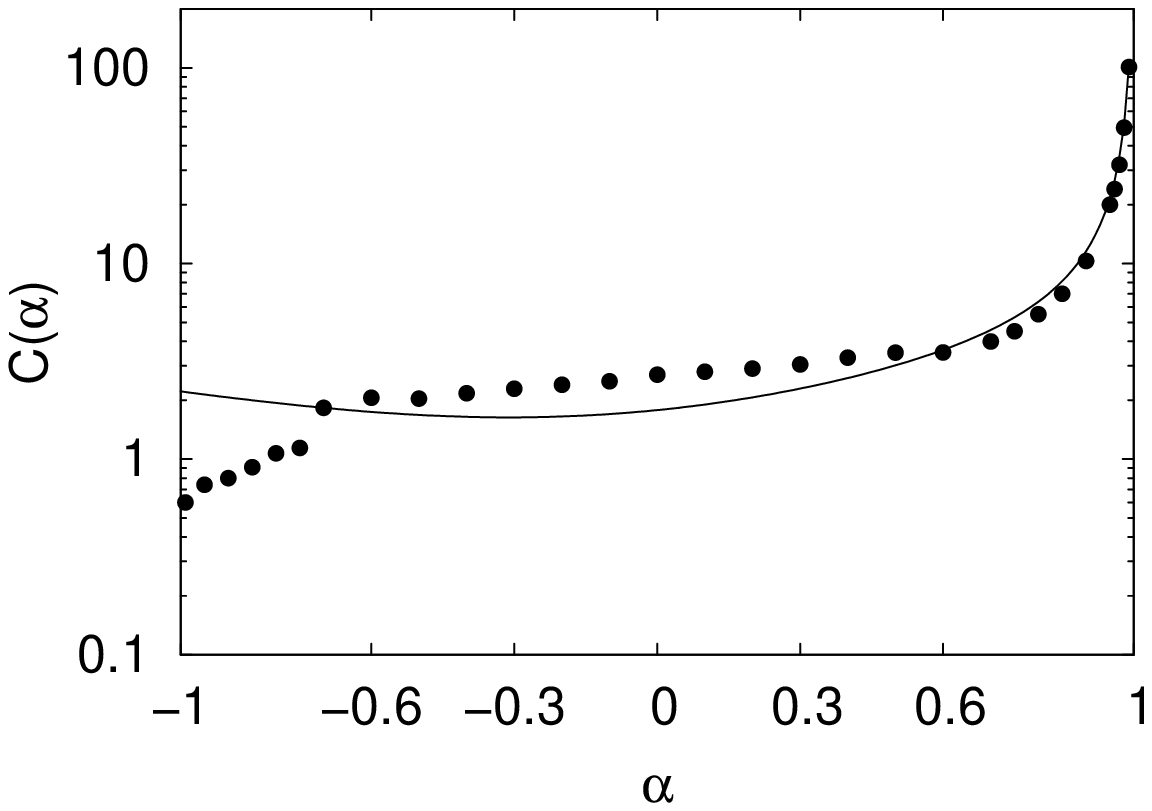}
 \caption{Left: $[a_2(\alpha,\beta)-a_{2,s}(\alpha)]/[a_2(\alpha,\beta_0)-a_{2,s}(\alpha)]$ for a two dimensional system as a function of $\beta_0(\beta-1)/[\beta(\beta_0-1)]$ for $\beta_0=10^{-2}$ (solid-black) and $\beta_0=10^2$ (dashed-blue), for $\alpha=0.99; 0.95; 0.85; 0.75; -0.60$ (from right to left), and for an initial gaussian distribution. Right: the exponent $C(\alpha)$ obtained from a best fit to a power law (symbols) and the the corresponding theoretical prediction of Eq. (\ref{eq:46}) (line).}
 \label{fig:3}
\end{figure}

As we did for the steady--state equation (\ref{eq:19}), in order to obtain an isotropic solution to Eq. (\ref{eq:41}), we assume that $\phi$ can be expanded in Sonine polynomials, as in Eq. (\ref{eq:20}), with the cumulants being functions of $\beta$. By multiplying Eq. (\ref{eq:41}) by an appropriate polynomial and integrating over $\mathbf c$, equations for the cumulants can be obtained. For example, by multiplying by $c^2$, we obtain 
\begin{eqnarray}
  \label{eq:42}
    && \frac{1}{4}\left\{\beta\mu_2[1|\phi]-\mu_2[\beta|\phi]\right\}\beta \frac{d}{d\beta}a_2(\beta) +\mu_2[\beta|\phi](1+a_2(\beta))-\mu_4[\beta|\phi]=0,
\end{eqnarray}
with
\begin{equation}
  \label{eq:43}
  \mu_4[\beta|\phi]=-\frac{1}{d(d+2)}\int d\mathbf c\ c^4J[\mathbf c,\beta|\phi].
\end{equation}
The equation for $a_2$ is coupled to the rest of the coefficients because of the functional dependence of $\mu_2$ and $\mu_4$ on $\phi$. However, if the latter quantities are expanded up to linear order in $a_2$ as
\begin{eqnarray}
  \label{eq:44}
    && \mu_2\simeq \mu_2^{(0)}+\mu_2^{(2)}a_2, \\
    && \mu_4\simeq \mu_4^{(0)}+\mu_4^{(2)}a_2,
\end{eqnarray}
with $\mu_i^{(j)}$ being known functions of $\alpha$ and $d$ given in the appendix, and we neglect contributions from cumulants of higher orders, then the solution to Eq. (\ref{eq:41}) can be written as
\begin{equation}
  \label{eq:45}
    a_2(\alpha,\beta) \simeq a_{2,s}(\alpha)+\left[a_2(\alpha,\beta_0)-a_{2,s}(\alpha)\right]\left[\frac{\beta_0(\beta-1)}{\beta(\beta_0-1)}\right]^{C(\alpha)},
\end{equation}
where $a_{2,s}(\alpha)$ is the steady--state value of the cumulant given by Eq. (\ref{eq:21}), $\beta_0$ is the initial value of $\beta$, and 
\begin{equation}
  \label{eq:46}
  C(\alpha)=4\left(\frac{\mu_4^{(1)}-\mu_2^{(1)}}{\mu_2^{(0)}}-1\right).
\end{equation} 
The exponent is $C\sim 2$ for $\alpha\le 0.5$ and diverges as $1/(1-\alpha)$ for $\alpha\sim 1$, see right plot of Fig. \ref{fig:3}. In order to be consistent, we should guarantee that $a_2$ is small, which can be controlled with an initial value close enough to the steady value, regardless the value of $\beta$. 

As a first application of Eq. (\ref{eq:46}), reconsider Eq. (\ref{eq:38}) that we rewrite as $\frac{d}{ds}\beta=F(\beta)$, with $F(\beta)\simeq \mu_2^{(0)}(1-\beta)+\mu_2^{(1)}\left[a_2(\alpha,\beta)-a_{2,s}(\alpha)\beta\right]$ at the first Sonine approximation. Since $C(\alpha)>1$, it is $\frac{d}{d\beta}F(1)\simeq -(\mu_2^{(0)}+\mu_2^{(1)}a_{2,s}(\alpha))<0$ implying that $\beta=1$ is a stable fixed point of (\ref{eq:38}) for all $\alpha\in[-1,1]$. That is, within the first Sonine approximation, the temperature and cumulant $a_2$ always reaches its steady--state values, provided we are close enough to $\beta=1$.

\begin{figure}[!h]
 \centering
 \includegraphics[width=.475\textwidth]{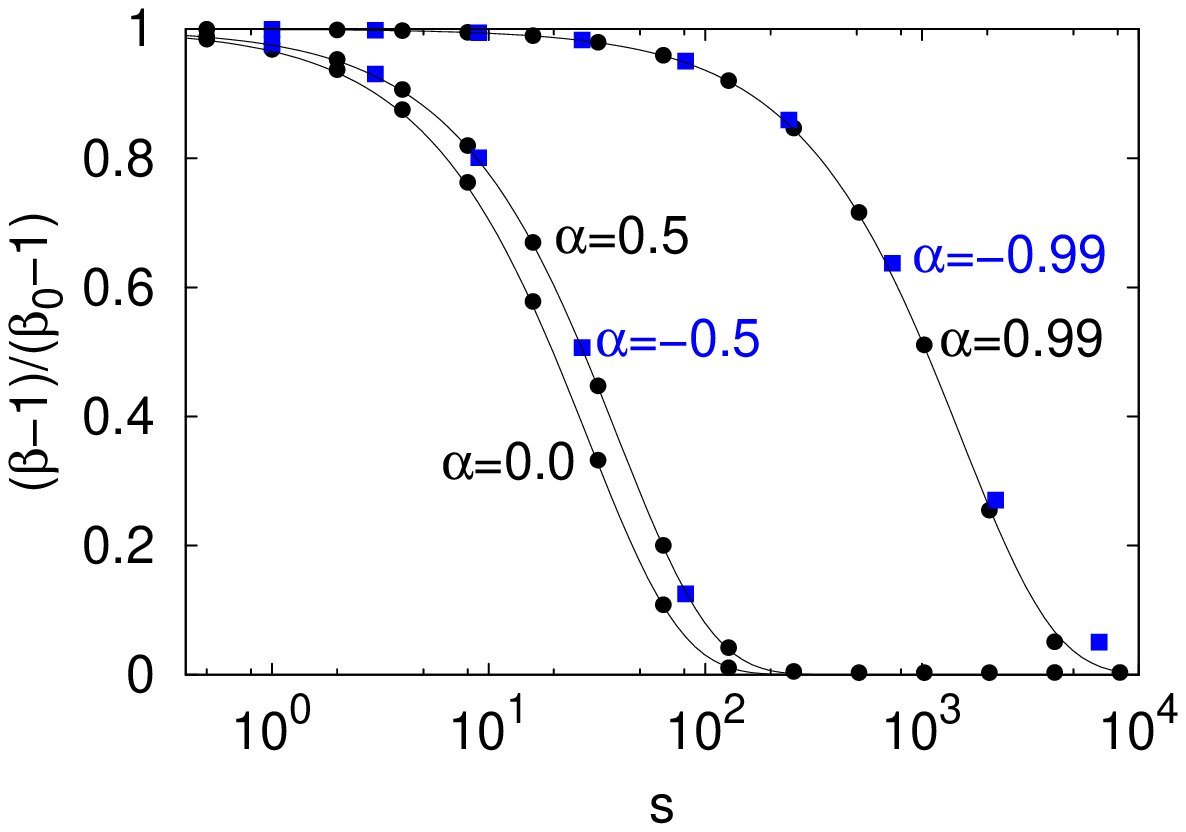}
 \includegraphics[width=.475\textwidth]{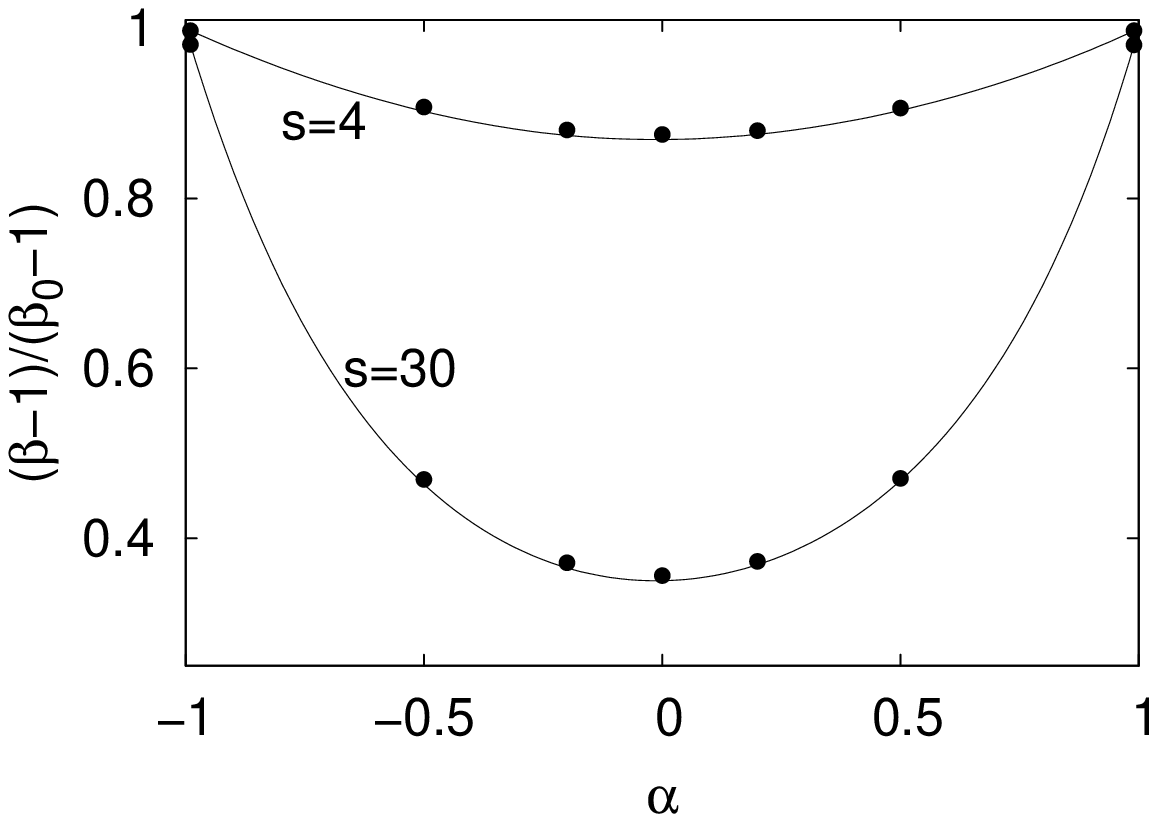}
 \caption{Time dependence of $\beta$ for different values of the parameters of the system.}
 \label{fig:5}
\end{figure}

As a second application, we compute the time dependence of $\beta$, and hence that of the temperature. In general, if we use expression (\ref{eq:45}) of the cumulant with Eq. (\ref{eq:38}),  the resulting equation for $\beta(s)$ turns out to be highly nonlinear. An important simplification occurs when $\beta$ is close to $1$, since it is $\mu_2[\beta|\phi]\simeq\mu_2[1|\phi]$, and the solutions $\beta(s)$ reads 
\begin{equation}
  \label{eq:47}
  \beta(s)\simeq 1+(\beta_0-1)e^{-\frac{1}{2}\zeta^*_ss},
\end{equation}
where we have used $\mu_2[1|\phi]=\frac{1}{2}\zeta^*_s$. The latter expression coincides, after using Eqs. (\ref{eq:39}) and (\ref{eq:17}), with Eq. (33) of \cite{brrumo04}, which in principle is only valid close enough to the steady state (see the comments on Fig. \ref{fig:5} bellow). 

\begin{figure}[!h]
 \centering
 \includegraphics[width=.475\textwidth]{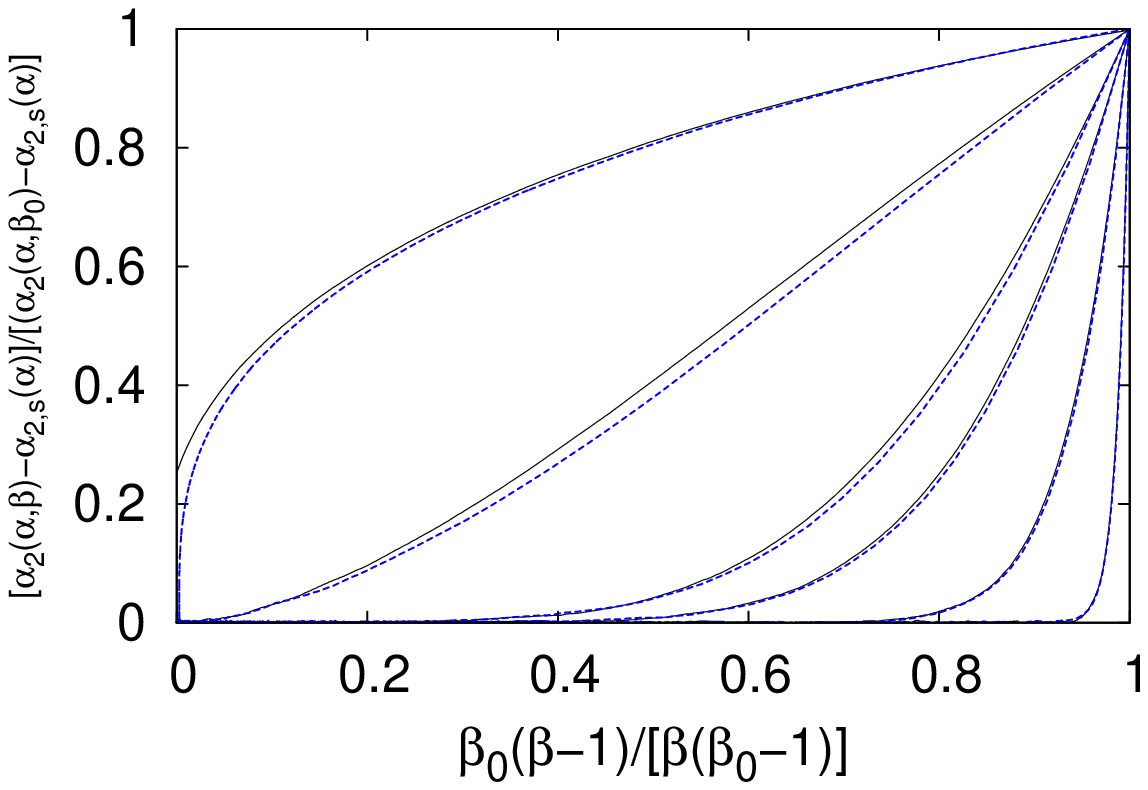}
 \includegraphics[width=.475\textwidth]{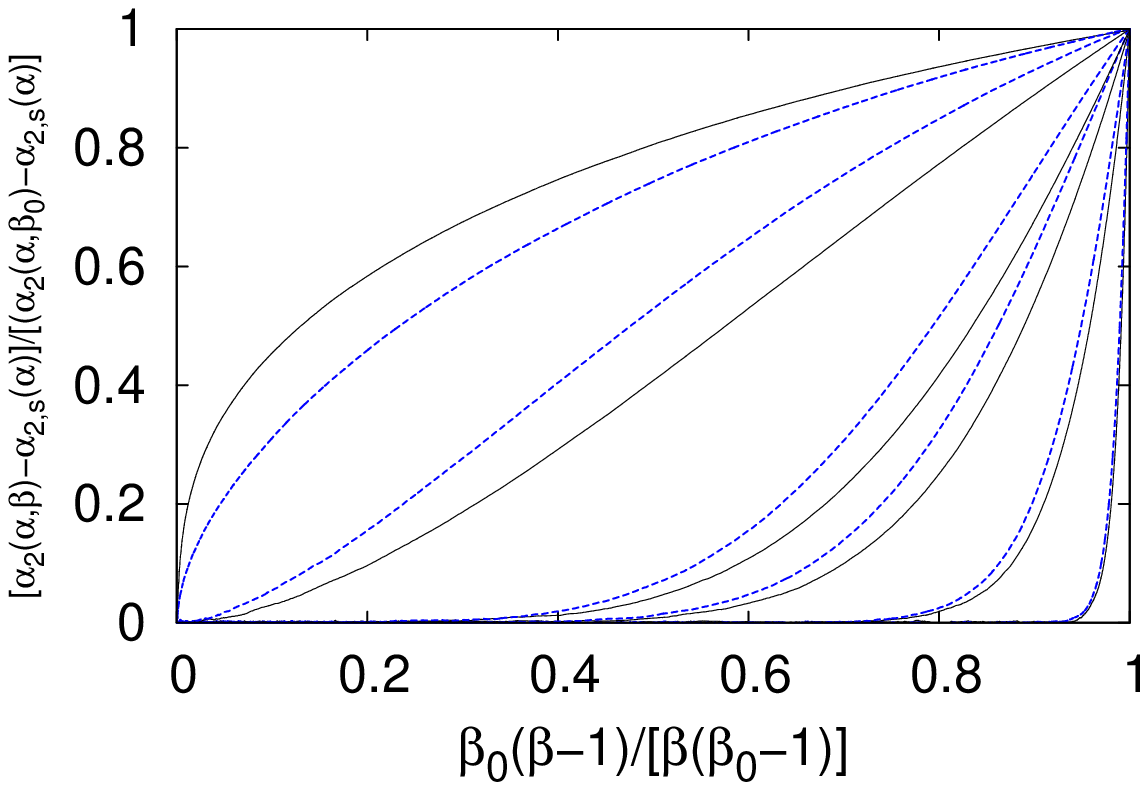}
 \caption{ $[a_2(\alpha,\beta)-a_{2,s}(\alpha)]/[a_2(\alpha,\beta_0)-a_{2,s}(\alpha)]$ for a two dimensional system as a function of $\beta_0(\beta-1)/[\beta(\beta_0-1)]$ for $\beta_0=10^{-2}$ (solid-black) and $\beta_0=10^2$ (dashed-blue), and for $\alpha=0.99; 0.95; 0.85; 0.75;-0.60;-0.95 $ (from right to left). Left: initial uniform distribution for which $a_2=-0.3$ and $a_{3}\simeq -0.29$. Right: initial distribution made of delta functions with $a_2\simeq -0.33$ and $a_{3}\simeq -0.44$.}
 \label{fig:4}
\end{figure}

Two predictions of the present section, that of Eqs. (\ref{eq:45})--(\ref{eq:47}), are compared against numerical simulations in Figs. \ref{fig:3} and \ref{fig:5}. On the one hand, Eq. (\ref{eq:45}) is confirmed by Fig. \ref{fig:3} for a range of values of $\alpha$ for which the first Sonine approximation is expected to work, namely the range provided by the analysis of the steady state, see Fig. \ref{fig:1}. For other values of $\alpha$, and if the initial cumulants are not small enough, the dependence of $\alpha_2$ on $\beta$ is also given by a law similar to Eq. (\ref{eq:45}), but with a different exponent $C(\alpha)$, as Figs. \ref{fig:3} and \ref{fig:4} show. More precisely, Fig. \ref{fig:4} suggests the same origin of the failure of Eq. (\ref{eq:45}) as that of Eq. (\ref{eq:21}), that is, the contribution of cumulants of higher order. This is because we observe a data collapse, similar to that predicted by Eq. (\ref{eq:45}), only if all initial values have the same cumulants (left plot of Fig. \ref{fig:4}) but this collapse is absent if the initial data have the same initial $a_2$ but different $a_{3}$ (right plot of Fig. \ref{fig:4}). On the other hand, the theoretical prediction of Eq. (\ref{eq:47}) is fully confirmed by Fig. \ref{eq:5} for all values of the parameters and initial conditions, meaning that the contribution of the cumulants can be neglected even if the first Sonine approximation fails.

\subsection{Relevance of the new scaling}

In order to clarify the relevance of the new scaling solution, Eq. (\ref{eq:35}), to the time evolution of the HCS, let us consider $e(s)$, a quantity proportional to the number of collisions per particle the system undergoes until time $s$. It can be defined such as $de=\frac{N}{L^d}\sigma^{d-1}\sqrt{\frac{2T}{m}}(\nu_0t)^{-1}dt$, where we have used that the granular temperature is $(\nu_0t)^{-2}T$. After using the changes of time variables at Eqs. (\ref{eq:10}) and (\ref{eq:39}), and using the generally valid Eq. (\ref{eq:47}), we have
\begin{equation}
  \label{eq:48}
  e(s)=\frac{2}{\zeta^*_s}\ln\left[\frac{\beta(\beta_0-1)}{\beta_0(\beta-1)}\right].
\end{equation}
This expression allows us to estimate the number of collisions for the temperature $\beta$ and the cumulant $a_2$ to relax to their steady--state values. For the first case, if $e_\beta$ is the value of $e$ for which $(\beta-1)/(\beta_0-1)=1/10$, we have
\begin{equation}
  \label{eq:49}
  e_\beta\simeq \frac{2}{\zeta_s^*}\ln\left(1+\frac{9}{\beta_0}\right) \sim \frac{2}{\zeta_s^*(\alpha)}.
\end{equation}
where we have assumed that $\beta_0\sim 1$, something to be kept in mind for the forthcoming discussion. Let $e_a$ be the value of $e$ for which $[a_2(\alpha,\beta)-a_{2,s}(\alpha)]/[a_2(\alpha,\beta_0)-a_{2,s}(\alpha)]=1/10$, then
\begin{equation}
  \label{eq:50}
  e_a\simeq \frac{2}{C\zeta^*_s}\ln10\sim \frac{2}{C(\alpha)\zeta^*_s(\alpha)},
\end{equation}
where we have used Eq. (\ref{eq:45}) for any $\alpha$ as an estimation for $a_2$. If we assume that the time relaxation of the cumulants of the higher orders are at least of order of $e_a$, then we have 
\begin{equation}
  \label{eq:51}
  e_a/e_\beta\sim \frac{1}{C(\alpha)}
\end{equation}
as an estimation of the number of collisions needed for the cumulants to relax in relation to that of the granular temperature. 

Let $e_k\sim 1$ be the number of collisions needed for a dilute elastic gas of hard spheres or disks to relax, an estimation of the so called kinetic stage of the evolution of the granular gas. Then, according to the results obtained so far, we can identify tree different behaviours of the granular gas depending on the value of the coefficient of normal restitution. For $\alpha\sim 1$, the time relaxation of the temperature is very big ($e_\beta \gg e_k$) but $e_a\sim e_k$ since $C\zeta_s^*\sim 1$. That is, for the relevant hydrodynamic time scales ($\gg e_r$) the cumulants take their steady--state values, and with a very good approximation $\phi(\mathbf c,\beta)\simeq \phi_s(\mathbf c)$. The situation seems to extend up to $\alpha\sim 0.7$.

For $|\alpha|\lesssim 0.7$ we also have $\phi(\mathbf c,\beta)\simeq \phi_s(\mathbf c)$, but now due to another reason. As it is apparent from Figs. \ref{fig:1} and \ref{fig:3}, for this range of $\alpha$'s it is $C\zeta_s^*\sim 1$ and $\zeta_s^*\sim 1$, so $e_\beta\sim e_k$ and $e_a\sim e_k$. That is, despite $e_\beta\sim e_a$, the system needs few collisions to reach $\beta=1$, and again for any relevant hydrodynamic time scales the HCS is at its steady sate. 

For $\alpha\lesssim -0.7$, the coefficient $C(\alpha)$ keeps of order one, while $\zeta_s^*$ becomes big, hence $e_\beta\sim e_a\gg e_k$. That is, the simplifications of the other cases does not hold, and we are intended to consider $\phi(\mathbf c,\beta)$ and $\beta=\beta(s)$.

\section{Discussion and conclusions \label{sec:5}}

In this work we have proposed an extension of the existing studies on the homogeneous cooling state of a granular gas along two complementary directions. On the one hand, the usual hard--sphere collision rule has been generalized by allowing the coefficient of normal restitution to take positive and negative values $\alpha\in[-1,1]$. This way, more complex collision processes have been modeled, as well as new situations of the granular gas, where the relative importance of dissipation and equilibration can be tuned, have been considered. Now, the elastic limit can be reached with $\alpha=1$ (hard, elastic spheres) and also with $\alpha=-1$ (ideal, collisionless gas). We have tried to realize to what extent the negative values of $\alpha$ modify the existing picture for $\alpha>0$. On the other hand, we also reconsidered the time evolution of the HCS, motivated by recent advances on the research of driven granular gases. The study has been carried out by means of theoretical and numerical (DSMC) approaches, both based on the kinetic description provided by the inelastic Boltzmann equation.

At the long--time limit, the relevant information about the dynamics of the HCS is encoded through the scaling distribution function $\phi_s(c)$, defined in Eqs. (\ref{eq:12}) and (\ref{eq:14}). We have studied their first two cumulants and their tails for all values of $\alpha\in[-1,1]$. Three different regions of the space of $\alpha$ can be identified. For $\alpha\ge 1/\sqrt{2}$, we recovered the known properties of $\phi_s$, namely it stays close to the gaussian distribution in the thermal region, hence its cumulants are very small, and their tails are exponential. The region $|\alpha|\le1/\sqrt{2}$ is characterized by an increase of the deviation of $\phi_s$ from the gaussian distribution, while the values of the cumulants and exponential tails being almost independent of the sign of $\alpha$. The latter means that, for this intermediate region, collisions have the same effect on the system, despite they are different. This is an example of two different collision rules, but conserving the number of particles and linear momentum, and being associated with the same energy dissipation, give rise to similar global behaviours of the system. Finally, for $\alpha<-1/\sqrt{2}$, we observe an important change of the shape of $\phi_s$ with respect to the other two regions: the cumulants take much bigger values and the distribution function becomes multimodal, meaning that its maxima now occur for velocities different from zero, provided $\alpha\lesssim -0.75$. Within this region, the limit $\alpha\sim -1$ has been studied in details. A simplification of the Boltzmann equation occurs, since collisions induce very small changes to the velocities. That is, not only is the energy dissipation very small (like for $\alpha\sim 1$) but also the randomization induced by collision is also very weak (in contrast to $\alpha\sim 1$). The new equation, which is of the Fokker--Plank type but with the diffusion and drift coefficients being functionals of $\phi_s$, provides a useful framework where the new phenomenology can be studied. Among many interesting results, we highlight that $\phi_s\sim 0$ for $c\sim 0$, the probability of finding a particle with velocity $\mathbf c$ being accumulated around $|\mathbf c|=c_m$ ($\sim 1$ for the two dimensional case). Moreover, the exponential tails of $\phi_s$ appear before new algebraic ones. To the best of our knowledge, such velocity (multimodal) distributions have never been observed so far. 

The second part of the work has been intended to the time evolution of the HCS for a wider time window. We have shown that a consistent solution to the Boltzmann equation requires a scaling distribution function to depend on two dimensionless parameters $\phi(\mathbf c,\beta)$, the dimensionless velocity already present at the long--time limit ($c$) and a new one that measures the deviation of the HCS from its long--time limit ($\beta$). We have provided explicit expressions for the first cumulant as a function of $\beta$, Eq. (\ref{eq:45}), as well as for the time dependence of the granular temperature, Eq. (\ref{eq:47}). The former turns out to be very accurate only if the cumulants are small enough, for any initial condition and for all relevant times (after a short transient of view collision per particle). The latter, on the contrary, is of general use. Despite the limitation of Eq. (\ref{eq:45}), the simulations shows that, in general, the second cumulant behaves in a similar way for any $\alpha\in[-1,1]$, which in turn represents an important support of the existence of $\phi(\mathbf c,\beta)$. 

The possibility of generalizing the results obtained for  $\phi(\mathbf c,\beta)$, Eqs. (\ref{eq:45}) and (\ref{eq:47}), to any value of $\alpha$, allowed us to discuss the actual importance of the new scaling solution in terms of the time the system takes to reach $\phi_s(\mathbf c)$. We have shown that for the relevant hydrodynamic time scales, the new scaling is only appreciable for $\alpha \le-1/\sqrt{2}$, provided the initial distribution is close enough to its long-time form. We find this as a very important result that gives support to the hydrodynamic description constructed from $\phi_s$. Nevertheless, we consider it necessary to consider $\phi(\mathbf c,\beta)$ if we were to wider the range of applicability of the aforementioned description. 

\ack
We acknowledge financial support from Ministerio de Economía y Competitividad (MINEICO) and Fondo Europeo de Desarrollo Regional (FEDER) under project ESOTECOS FIS2015-63628-C2-1-R. 

\appendix

\section{Coefficients $\mu_i^{(j)}$}

The coefficients $\mu_i^{(j)}$ can be obtained from \cite{noer98} after a slight change of the notation
\begin{eqnarray}
  \label{eq:a1}
  && \mu_2^{(0)}=\frac{\sqrt{2\pi}(1-\alpha^2)}{2d\Gamma\!\left(\frac{d}{2}\right)}, \\
  \label{eq:a2}
  && \mu_2^{(2)}=\frac{3}{16}\mu_2^{(0)}, \\
  \label{eq:a3}
  && \mu_4^{(0)}=\frac{d+\frac32+\alpha^2}{d+2}\mu_2^{(0)}, \\
  \label{eq:a4}
  && \mu_4^{(2)}=\frac{\mu_2^{(0)}}{(d+2)\left[\frac{3}{32}(10d+39+10\alpha^2)+\frac{d-1}{1-\alpha}\right]}.
\end{eqnarray}

\end{document}